\documentclass[twocolumn,amssymb, amsmath,aps,prb,superscriptaddress]{revtex4-1}
\usepackage{txfonts}
\usepackage{bm}
\usepackage{graphicx}
\usepackage{braket}
\usepackage{here}
\usepackage{color}
\usepackage{breqn}

\makeatletter
\let\cat@comma@active\@empty
\makeatother

\begin{document}

\title{Effect of on-site Coulomb repulsion on ferromagnetic fluctuations in heavily overdoped cuprates} 

\author{Shingo Teranishi}%
\affiliation{Department of Materials Engineering Science, Graduate School of Engineering Science, Osaka University, 1-3 Machikaneyama-cho, Toyonaka, Osaka 560-8531, Japan}%

\author{Kazutaka Nishiguchi}%
\affiliation{Department of Physics, Graduate School of Science, Osaka University, 1-3 Machikaneyama-cho, Toyonaka, Osaka 560-8531, Japan}%

\author{Seiji Yunoki}%
\affiliation{Computational Materials Science Research Team, RIKEN Center for Computational Science (R-CCS), Kobe, Hyogo 650-0047, Japan}%
\affiliation{Computational Condensed Matter Physics Laboratory, RIKEN Cluster for Pioneering Research (CPR), Wako, Saitama 351-0198, Japan}%
\affiliation{Computational Quantum Matter Research Team, RIKEN, Center for Emergent Matter Science (CEMS), Wako, Saitama 351-0198, Japan}%

\author{Koichi Kusakabe}%
\affiliation{Department of Materials Engineering Science, Graduate School of Engineering Science, Osaka University, 1-3 Machikaneyama-cho, Toyonaka, Osaka 560-8531, Japan}%

\date{\today}

\begin{abstract}
We theoretically study ferromagnetic (FM) fluctuations that are experimentally observed in the heavily overdoped region of cuprate superconductors. To explore the origin of FM fluctuations, we evaluate the spin susceptibilities of a single-band Hubbard model within the fluctuation exchange approximation. Model parameters are derived using the Wannierization technique and the constrained random phase approximation method based on the maximally localized Wannier functions. The constrained random phase approximation calculations reveal that the on-site Coulomb interaction decreases with an increase in hole doping. By taking this reduction of the on-site Coulomb interaction into account, the emergence of FM fluctuations in heavily overdoped cuprates can be explained.
\end{abstract}

\maketitle

\section{Introduction}
Several studies have demonstrated the presence of ferromagnetic (FM) fluctuations in the heavily overdoped regime of cuprates. Experimentally, Kopp et al.~\cite{FMkopp} discussed the magnetic susceptibility in non-superconducting heavily overdoped Tl-2201 and argued that the competition between FM fluctuations and superconductivity is responsible for the termination of the superconducting dome in the overdoped regime.
Kurashima et al.~\cite{FMkurashima} reported the characteristic behaviors of metals with FM fluctuations~\cite{scr_1,scr_2}.

In terms of the theoretical aspect, a Monte Carlo method shows a transition from antiferromagnetic (AFM) to FM spin correlations with an increase in hole doping~\cite{jia2014persistent}. The existence of an FM ground state is suggested by the two-particle self-consistent (TPSC) approach by investigating the two-dimensional Hubbard model on a square lattice at Van Hove densities~\cite{FM_TPSC_Tremblay}. The TPSC approach also shows the existence of an incommensurate FM peak near $\bm{q}=0$ for several fillings~\cite{OguraTPSC}. Using dynamic cluster approximation calculations, Mayer et al.~\cite{maierscalapino2020,PhysRevResearch.2.033132} reported that FM fluctuations arise naturally in a spin-fluctuation framework. According to electronic band calculations, weak ferromagnetism can appear locally around the clusters of high Ba concentration in supercells of La$_{(2-x)}$Ba$_x$CuO$_4$~\cite{Barbiellini}.

Motivated by these studies, we discuss the origin of FM fluctuations in heavily overdoped cuprates by considering electron--electron correlation effects. Our results obtained using a single-band model in an overdoped regime suggest that spin fluctuations around $\bm q = (0, 0)$ become larger than those around $\bm q = (\pi,\pi)$ when the on-site Coulomb interaction $U$ is in an appropriate range of $3 \lesssim U \lesssim 4.5$ eV. This range is consistent with our evaluation of the screened on-site Coulomb interaction in overdoped cuprates using constrained random phase approximation (constrained-RPA). Here, we propose that it is necessary to consider filling dependence of on-site Coulomb interaction to fully understand FM fluctuations in heavily overdoped cuprates. Because the on-site Coulomb interaction has been treated as a constant in many previous studies, our scenario can provide a new perspective, not only for FM fluctuations in overdoped cuprates but also for the origin of superconductivity.

In this study, we calculate spin susceptibilities of the single-band Hubbard model on a square lattice by applying fluctuation exchange approximation (FLEX)~\cite{FLEX_1,FLEX_2} because FLEX is suitable for analyzing systems with strong spin fluctuations. In addition, we evaluate the electronic structures given by the generalized gradient approximation of density functional theory (DFT) in several cuprates to compare cuprates with several hole concentrations. To evaluate the strength of the on-site Coulomb interaction for several hole concentrations, we apply constrained-RPA.

\section{Methods}
\subsection{model calculation}
First, we introduce FLEX. A single-band Hubbard Hamiltonian is represented as
\begin{align}
H=\sum_{ij\sigma}\left( t_{ij}c^\dag_{i\sigma}c_{j\sigma}+H.c.\right)+U\sum_in_{i\uparrow}n_{i\downarrow},
\end{align}
where $c^\dag_{i\sigma}$ ($c_{i\sigma}$) represents the electron creation (annihilation) operator at site $i$ on a square lattice and spin $\sigma$. $n_{i,\sigma}$ represents the particle number operator at site $i$ and spin $\sigma$. The Hubbard model in the non-interacting limit ($U=0$) corresponds to a simple tight-binding model. The energy dispersion $\xi_{\bm k}$ of the tight-binding model is
\begin{dmath}
\xi_{\bm k} = -2t(\cos{k_x}+\cos{k_y}) + 4t'\cos{k_x}\cos{k_y} -2t''(\cos{2k_x}+\cos{2k_y})-\mu.
\end{dmath}
where $\mu$ represents the chemical potential, and $t$,$ t'$, and $t''$ represent the nearest-, second-,
and third-neighbor hoppings, respectively.

The interacting Green's function $G(\bm k,i\varepsilon_n) $ is obtained as
\begin{equation}
G(\bm k,i\varepsilon_n) = \left[G_0(\bm k,i\varepsilon_n)^{-1}- \Sigma(\bm k,i\varepsilon_n)\right]^{-1},
\label{dyson}
\end{equation}
where $G_0(\bm k,i\varepsilon_n)$ is the non-interacting Green's function,
\begin{equation}
G_0(\bm k,i\varepsilon_n) = \frac{1}{i\varepsilon_n-\xi_{\bm k}}.
\end{equation}

The longitudinal and
transverse spin susceptibilities are defined respectively as 
\begin{equation}
\chi^{zz}_{\mathrm{spin}}(\bm q,i
\omega_m)=\int_0^\beta d\tau e^{i\omega_m\tau}\left<S_{\bm q}^z(\tau)S_{\bm{-q}}^z(0)\right>
\end{equation}
and
\begin{equation}
\chi^{+-}_{\mathrm{spin}}(\bm q,i\omega_m)=\int_0^\beta d\tau e^{i\omega_m\tau}\left<S_{\bm q}^+(\tau)S_{\bm{-q}}^-(0)\right>.
\end{equation}
Spin susceptibility within FLEX is obtained as 
\begin{equation}
\chi_{\mathrm{spin}}(\bm q,i\omega_m)=\frac{\chi_0(\bm q,i\omega_m)}{1-U\chi_0(\bm q,i\omega_m)},
\end{equation}
where
\begin{equation}
\chi_0(\bm q,i\omega_m) \equiv -\frac{1}{N\beta}
\sum_{\bm k, n}G_0(\bm k,i\varepsilon_n
)G_0(\bm k + \bm q,i\omega_m+i\varepsilon_n)
\end{equation}
is the irreducible susceptibility. 

The self-energy is obtained as 
\begin{equation}
\Sigma(\bm k,i\varepsilon_n)=\frac{1}{N\beta}\sum_{\bm q,m}V^{\Sigma}(\bm q, i\omega_m)G(\bm k -\bm q, i\varepsilon_n-i\omega_m)
\label{self}
\end{equation}
with
\begin{dmath}
\nonumber
V^{\Sigma}(\bm q,i\omega_m)=U^2\left[\frac{3}{2}\left(\frac{\chi_0(\bm q,i\omega_m)}{1-U\chi_0(\bm q,i\omega_m)}\right)\\+\frac{1}{2}\left(\frac{\chi_0(\bm q,i\omega_m)}{1+U\chi_0(\bm q,i\omega_m)}\right)-\chi_0(\bm q,i\omega_m)\right],
\end{dmath}
Here, $i\varepsilon_n=i2\pi(n-1)k_{\mathrm B}T$ is the Matsubara frequency for fermions, $i\omega_m=i2\pi mk_{\mathrm B}T$ is the Matsubara frequency for bosons, 
$N$ is the number of sites, and $k_{\mathrm B}T=1/\beta$ is the temperature. 
In FLEX, the interacting Green's function $G(\bm k,i\varepsilon_n) $ is self-consistently determined by solving these equations. The chemical potential $\mu$ has to be fixed self-consistently for a given electron concentration. 

To discuss the strength of superconductivity, we solve the linearized Eliashberg equation
\begin{equation}
 \lambda\Delta(\bm k,i\varepsilon_n)=-\frac{1}{N\beta}\sum_{\bm k',m} V^s(\bm k-\bm k',i\varepsilon_n-i\varepsilon_n')|G(\bm k',i\varepsilon_n')|^2\Delta(\bm k', i\varepsilon_n').
\end{equation}
Here, $\Delta(\bm k,i\varepsilon_n)$ is a gap function, $G(\bm q,i\varepsilon_n')$ is the dressed Green's function, and $\lambda$ is the eigenvalue of the Eliashberg equation, where $\lambda=1$ corresponds to $T=T_c$; therefore, $\lambda$ serves as a measure of the strength of superconductivity. $V^s(\bm k-\bm q,i\varepsilon_n-i\varepsilon_n')$ is the effective interaction for a spin singlet pairing, which is represented as 
\begin{dmath}
V^s(\bm q,i\omega_m)=U^2\left[\frac{3}{2}\left(\frac{\chi_0(\bm q,i\omega_m)}{1-U\chi_0(\bm q,i\omega_m)}\right)-\frac{1}{2}\left(\frac{\chi_0(\bm q,i\omega_m)}{1+U\chi_0(\bm q,i\omega_m)}\right)\right]+U,
\end{dmath}
In our calculations, we take $N=256^2$ sites with 2048 Matsubara frequencies, and $k_{\mathrm B}T=0.01$~eV. We consider several filling factors of $n=0.85,~0.8,~0.75~(25\%~\mathrm{doped}),~0.7,~0.65,~0.6~(40\%~\mathrm{doped})$ and when $n=1.0$, the system is half-filling. Hopping parameters for the $3\mathrm{d}_{x^2-y^2}$ bands in this research are $(|t'/t|,|t''/t|)=(0.2266,0,2111)$, and we set $|t|=1$ for simplicity.
As the angle-resolved photoemission spectroscopy (ARPES) experiment of HgBa$_2$CuO$_{4+\delta}$~\cite{Hg1201_ARPES} shows a Fermi surface that is similar to that of our model, we argue that this model is not too far from the reality. The density of states of the tight-binding model is calculated using the tetrahedron method~\cite{tetra_origin_1,tetra_origin_2}.

\subsection{First principle calculations}

\begin{figure}[ht]
 \centering
  \includegraphics[keepaspectratio, scale=0.35]{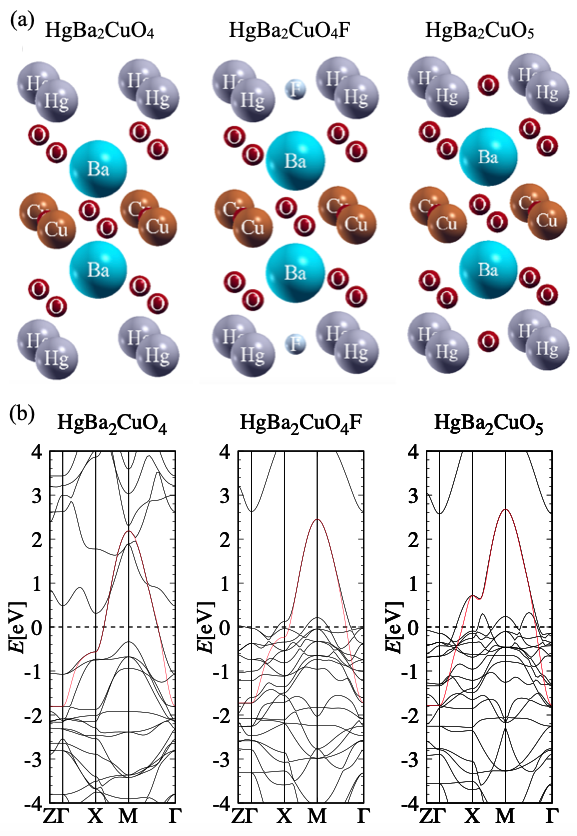}
\caption{(a) Atomic structures of HgBa$_2$CuO$_4$, HgBa$_2$CuO$_4$F, and HgBa$_2$CuO$_5$. (b) Band structures of HgBa$_2$CuO$_4$, HgBa$_2$CuO$_4$F, and HgBa$_2$CuO$_5$. Wannier interpolated bands are plotted with a red line. Here, the Fermi level is set to 0.\label{fig:str_band}}
\end{figure}
\begin{figure*}[ht]
    \includegraphics[keepaspectratio, scale=0.5]{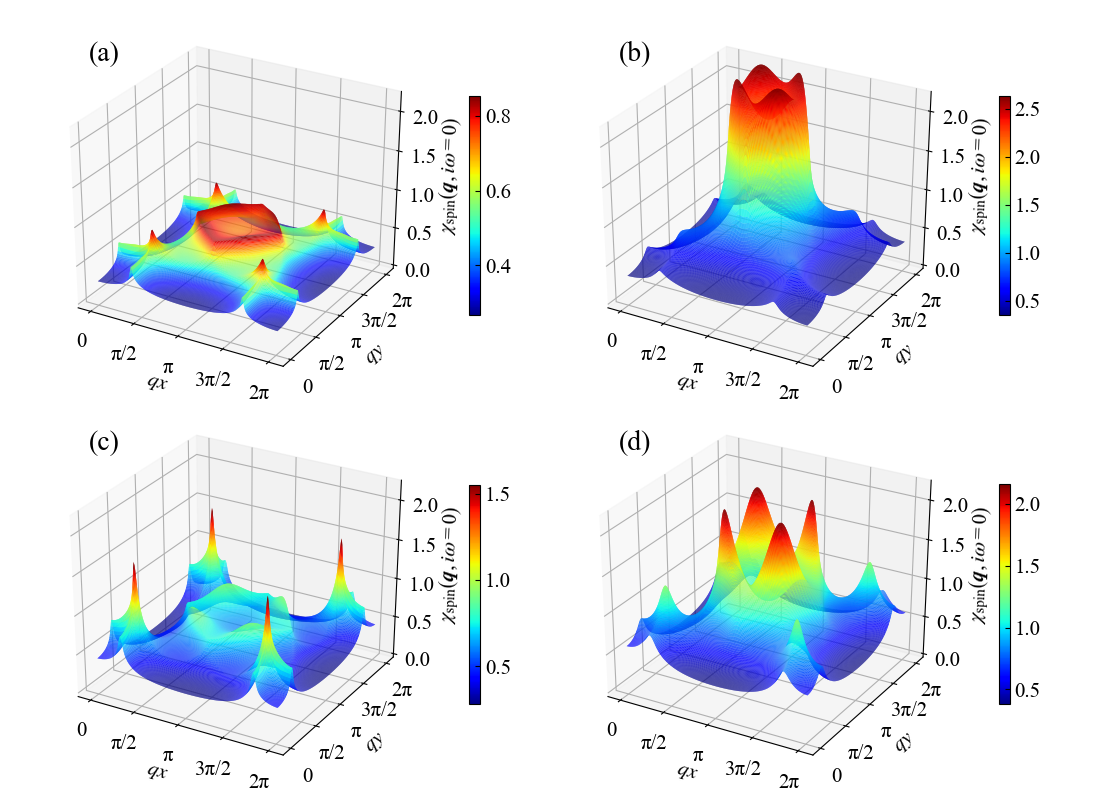}
        \caption{Spin susceptibility $\chi_{\mathrm{spin}}(\bm q,i\omega_m=0)$ calculated using FLEX. (a) $U=3.0$~eV and $n=0.85$, (b) $U=6.0$~eV and $n=0.85$, (c) $U=3.0$~eV and $n=0.7$, and (d) $U=6.0$~eV and $n=0.7$. Here, $k_{\mathrm B}T = 0.01$~eV.~\label{fig:spinU3_6_001_035_0425}}
\end{figure*}
In this study, we analyze several Hg-compounds that have different doping concentrations, such as HgBa$_2$CuO$_4$~\cite{Hg1201-first}, HgBa$_2$CuO$_4$F, and HgBa$_2$CuO$_5$. The crystal structures of these systems are shown in Fig.~\ref{fig:str_band}(a). For HgBa$_2$CuO$_4$F and HgBa$_2$CuO$_5$, there are several studies that consider the relevant crystal phases~\cite{Hg1201_F_DFT,Hg1201_F_neutron,HgBa2CuO5_DFT}. Although the FM fluctuations are not observed in Hg-compounds, Hg-compounds are considered to be a model cuprate because these have the highest $T_{\mathrm c}$ among known single-layer cuprates~\cite{Sakakibara2012}. In addition, a related triple-layer Hg-compound, Hg-1223, shows the highest $T_{\mathrm c}$ among all cuprates. From a practical aspect, it has a simple crystal structure; hence, it is easy to calculate.

Now, let us discuss nominal filling factors in the CuO$_2$ planes of these Hg-compounds. A nominal filling factor in the CuO$_2$ plane can be obtained by the rule for the ionization valence of a noble metal, an alkaline earth metal ion, and oxygen. Assuming $\rm{Hg}^{+2},\rm{Ba}^{+2}$, and $\rm{O}^{��'2}$, the formal valences of Cu for these compounds are +2 (half-filling) for HgBa$_2$CuO$_4$, +3 (one hole doped per unit cell) for HgBa$_2$CuO$_4$F, and +4 (two hole doped per unit cell) for HgBa$_2$CuO$_5$. One or two holes doped per unit cell is too overdoped for cuprates; however, it can be considered that those structures can capture properties in the overdoped limit of cuprates.

To obtain the electronic structures of several cuprates, we adopt reliable DFT codes~\cite{QE-2009,QE-2017} in our calculations. Then, we can obtain the hopping parameters among $3\mathrm{d}_{x^2-y^2}$ Wannier interpolated bands using the Wannierization technique~\cite{wannier_ent,wannier_loc,wannier90}. Table \ref{table:hop} shows the evaluated hopping parameters for the 3d$_{x^2-y^2}$ band in HgBa$_2$CuO$_4$, HgBa$_2$CuO$_4$F, and HgBa$_2$CuO$_5$. In Fig.\ref{fig:str_band}(b), we show the obtained band structures and Wannier interpolated bands. Here, the filling factors of the single-band tight-binding models via Wannierization at $k_{\mathrm B}T = 0.01$ eV are $n=0.9481$(HgBa$_2$CuO$_4$), $n=0.8250$(HgBa$_2$CuO$_4$F), and $n=0.5949$(HgBa$_2$CuO$_5$), respectively. Therefore, these crystal structures reflect different doping rates in the CuO$_2$ planes.
\begin{table}[h]
\caption{Hopping parameters for the 3d$_{x^2-y^2}$ band in HgBa$_2$CuO$_4$, HgBa$_2$CuO$_4$F, and HgBa$_2$CuO$_5$.\label{table:hop}}
  \small
  \begin{tabular}{lccc}
\hline
& HgBa$_2$CuO$_4$&HgBa$_2$CuO$_4$F&HgBa$_2$CuO$_5$\\
    \hline \hline
     $t$[eV]&-0.450&-0.500&-0.516\\ 
    \hline
     $t'$[eV]&0.102&0.073&0.029\\ 
     \hline
     $t''$[eV]&-0.095&-0.084&-0.061\\ 
     \hline
  \end{tabular}
\end{table} 
\subsection{constrained-RPA}
We introduce constrained-RPA, using which we can calculate an effective screened on-site Coulomb interaction\cite{cRPA}. In this method, we divide polarization ($P$) into two contributions. 
One is from the transitions among target bands ($P_d$), and the other is from the other transitions ($P_r$). The screened interaction $W$ at the RPA level is given by
 \begin{eqnarray}
 W&=&[1-vP]^{-1}v=[1-W_rP_d]^{-1}W_r, \label{scr_Coulomb_cRPA}
 \end{eqnarray}
where $v(\bm q)=4\pi/\Omega\bm |q|^2$ is the bare Coulomb interaction. 
In Eq.(\ref{scr_Coulomb_cRPA}), we have defined a screened interaction $W_r$ that does not take the polarization from 3d--3d transitions into account:
\begin{align}
W_r=[1-vP_r]^{-1}v, 
\label{eq:c-rpa}
\end{align}
which does not include the polarization from $3d$--$3d$ transitions.
Static screened Coulomb interactions in the Wannier basis are expressed as
\begin{align}
V_{ij}=\int d\bm r\int d\bm r'\phi^*_i(\bm r)\phi_i(\bm r) W_r(\bm r,\bm r')\phi^*_j(\bm r')\phi_j(\bm r').
\end{align}
Here, $i$ and $j$ are the indices of the Wannier orbitals. The Coulomb interaction 
$W_r(\bm r,\bm r')$ screened by $P_r$ is rewritten in a symmetric form, 
\begin{align}
W_r(\bm r, \bm r')=\frac{4\pi}{\Omega}\sum_{\bm{{qGG'}}}\frac{e^{-\mathrm{i}(\bm{{q+G}})\bm{{r}}}}{|\bm{{q+G}}|}\epsilon^{-1}_{c\bm{{GG'}}}(\bm{{q}})\frac{e^{-\mathrm{i}(\bm{{q+G'}})\bm{\mathrm{r'}}}}{|\bm{{q+G'}}|},
\end{align}
where $\Omega$ is the crystal volume, $\epsilon^{-1}_{c\bm{{GG'}}}(\bm{{q}})$ is the inverse dielectric matrix, $\bm q$ is a wave vector in the first Brillouin zone, and $\bm G$ is a reciprocal lattice vector. 
The dielectric matrix is expressed as 
\begin{align}
\epsilon_{c\bm{{GG'}}}(\bm q) = \delta_{\bm{{GG'}}}-\nu(\bm{q+G})\chi_{c\bm{{GG'}}}(\bm q), 
\end{align}
where the polarization matrix $\chi_{c\bm{{GG'}}}(\bm q)$ in constrained-RPA is given by
\begin{align}
\nonumber
\chi_{c\bm{{GG'}}}(\bm q)=\sum_{\bm k}\sum_{\alpha\beta}\bra{\psi_{\alpha\bm{k+q}}}e^{-\mathrm{i}(\bm{q+G})\bm{r}}\ket{\psi_{\beta\bm k}}\\
\times\bra{\psi_{\beta\bm{k}}}e^{-\mathrm{i}(\bm{q+G'})\bm{r}}\ket{\psi_{\alpha\bm {k+q}}}\frac{f_{\alpha\bm{k+q}}-f_{\beta\bm k}}{E_{\alpha\bm{k+q}}-E_{\beta\bm k}}.
\end{align}
Here, $|\psi_{\alpha\bm k}\rangle$ is a Bloch state with  
energy $E_{\alpha\bm k}$,  
and $f_{\alpha\bm k}$ is its occupancy. 
$\alpha,\beta$ stand for the bands that do not include $3d$--$3d$ band transitions. Explicitly, $f_{\alpha\bm k}=1$ for valence bands, 
and $f_{\alpha\bm k}=0$ for conduction bands.

To treat entangled bands in constrained-RPA, several approaches have been proposed~\cite{c-rpa_pd2,c-rpa_pd1}. In this study, $P_d$ is constructed by summing over all the transitions in the Lehmann representation multiplied by the probabilities of transitions that take place within the $d$ subspace~\cite{c-rpa_pd3}. To perform the calculation, we use the RESPACK code~\cite{respack1,respack2,respack3,respack4,respack5}.

During calculations, the wave function and charge density are expanded in plane waves with the cutoff energies of (100, 400) [Ry], and the cutoff energy for polarization functions is 10 [Ry]. We use the $8\times8\times8~k-$point mesh in the first Brillouin zone and 
take 100 bands into account in all calculations. The unit cell of all the compounds is optimized in the simulation, where the pressure is controlled using a criterion for each diagonal element of the stress tensor being less than 0.5 [kbar]. The internal atomic structures are optimized using a criterion that the summation of the absolute values of force vector elements becomes smaller than
$1.0\times10^{-8}$ [Ry/a.u.].

\section{Results}
\begin{figure}[ht]
\centering
    \includegraphics[keepaspectratio, width=85mm]{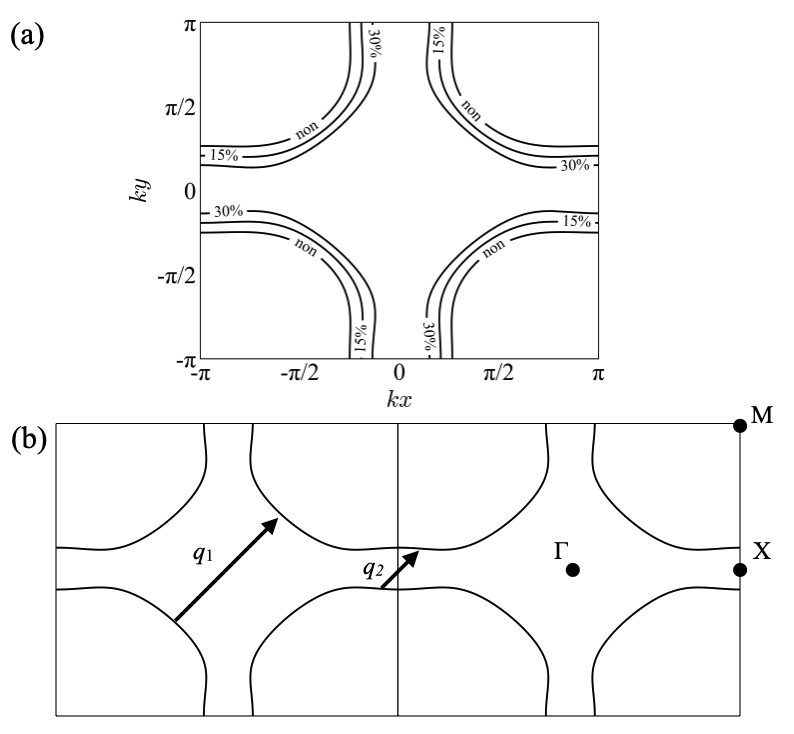}
        \caption{(a) Fermi surfaces of the tight-binding model for three fillings: non-doped case, 15\%-doped case ($n=0.85$), and 30\%-doped case ($n=0.7$). (b) Schematic figure of the Fermi surface with nesting vectors $\bm q_1$ and $\bm q_2$.\label{fig:Fermi_con_sch}}
\end{figure}
\begin{figure}[ht]
    \includegraphics[keepaspectratio, width=85mm]{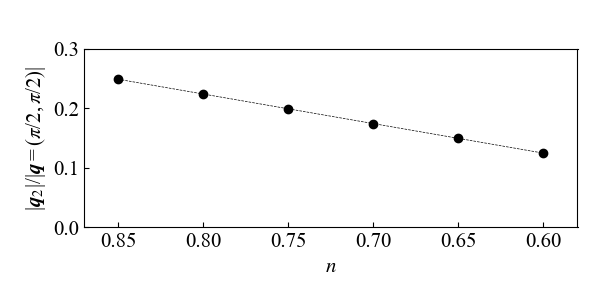}
        \caption{Change in $|\bm q_2|/|\bm q=(\pi/2,\pi/2)|$ as a function of electron concentration $n$. Here, we set $U=3$.\label{fig:q2}}
\end{figure}
\begin{figure}[ht]
 \centering
  \includegraphics[keepaspectratio, scale=0.5]{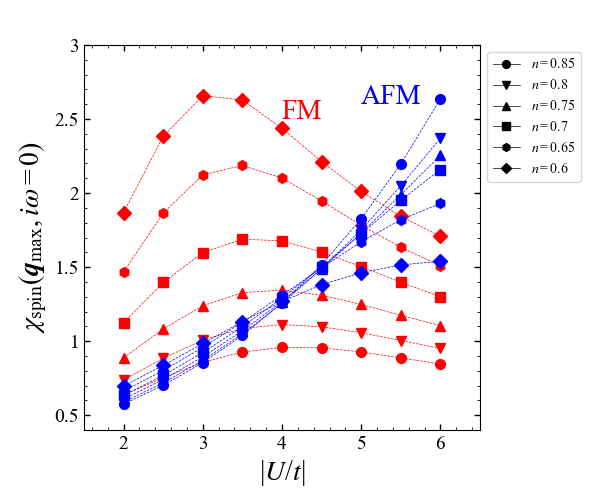}
\caption{Change in the maximum values of incommensurate FM fluctuations and AFM fluctuations as a function of on-site Coulomb interactions ($U$). Here, $k_{\mathrm B}T = 0.01$ eV.\label{fig:uvspeak}}
\end{figure}
\begin{figure}[ht]
    \includegraphics[keepaspectratio, width=90
    mm]{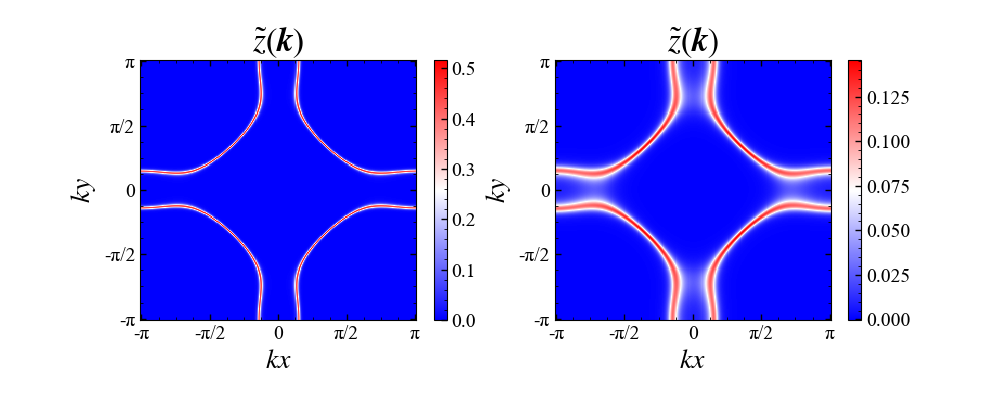}
        \caption{Average spectral function. We set $U=3.0$ eV in the left figure and $U=6.0$ eV in the right figure. Here, $n=0.7$ and $k_{\mathrm B}T = 0.01$~eV.
        \label{fig:z_U3_6_001_035_v2}}
\end{figure}
 \begin{figure}[ht]
    \includegraphics[keepaspectratio,width=85mm]{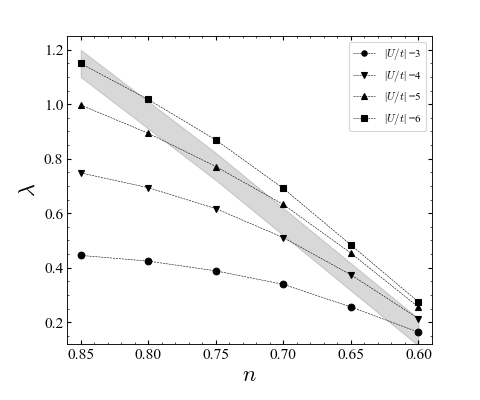}
         \caption{Eigenvalue $\lambda$ of the linearized Eliashberg equation with different amplitudes of $U$ in the single-band Hubbard model at $k_{\mathrm B}T = 0.01$~eV. Gray shaded line indicates a possible change in $\lambda$s due to the reduction in $U$ with hole doping.}\label{fig:lambda}
 \end{figure}

Figure~\ref{fig:spinU3_6_001_035_0425} shows the spin susceptibilities evaluated by applying FLEX for $U = 3$~eV and $U = 6$~eV at two doping cases. The Fermi surfaces of the tight-binding model in the optimal-doped case ($n=0.85$) and in the overdoped case ($n=0.7$) are shown in Fig.~\ref{fig:Fermi_con_sch}(a). The schematic Fermi surfaces of the tight-binding model with nesting wave vectors $\bm q_1$ and $\bm q_2$ are also shown in Fig.~\ref{fig:Fermi_con_sch}(b). Change in $|\bm q_2|$ with the electron concentrations is shown in Fig.~\ref{fig:q2}.

One can see that the FM peaks are located in $0 \leq q_x,q_y \leq \pi/2$, and the AFM incommensurate peaks are located in $\pi/2 < q_x,q_y < \pi$. Both peaks are associated with the nesting of the Fermi surface, as shown in Fig.~\ref{fig:Fermi_con_sch}(b). Specifically, the FM nesting vector $\bm q_2$ is located at small momentum around $\bm q = 0$; thus, the spin susceptibilities obtained by applying FLEX show the peaks around the wave vector corresponding to $\bm q_2$. With hole doping, the FM nesting vector $\bm q_2$ moves close to $(0, 0)$, as shown in Fig.~\ref{fig:q2}. 

Table~\ref{table:U} shows the evaluated on-site Coulomb interaction obtained by applying constrained-RPA. One can see that the value of bare on-site Coulomb interaction $U_{\mathrm{bare}}$ in several Hg-compounds is similar. However, interestingly, the screened on-site Coulomb interaction in Hg-compounds reduces with an increase in hole doping; this indicates that reduction in $U$ is mainly due to the change in the dielectric matrix $\epsilon^{-1}_c$. This tendency is consistent with the previous study on HgBa$_2$CuO$_4$ and TlBa$_2$CuO$_5$\cite{cRPAteranishi}.
Figure~\ref{fig:uvspeak} shows changes in the maximum values of incommensurate AFM fluctuation and FM fluctuation as a function of strength of on-site Coulomb interactions ($U$) for several electron concentrations $n$. While the incommensurate AFM fluctuation monotonically increases, the incommensurate FM fluctuation increases once, and then it decreases with an increase in $U$. For example, in the case of $n=0.7$, it increases from $U = 2$ eV to $U \simeq 4.0$ eV; then, it decreases from $U \simeq 4$ eV to $U = 6$ eV. We notice that the incommensurate FM fluctuation becomes larger than the incommensurate AFM fluctuation at $U \lesssim 4.5$ eV.

\begin{table}[h]
  \caption{Screened on-site Coulomb interaction evaluated by constrained-RPA in HgBa$_2$CuO$_4$, HgBa$_2$CuO$_4$F, and HgBa$_2$CuO$_5$.\label{table:U}}
  \small
  \begin{tabular}{lccc}
\hline
& HgBa$_2$CuO$_4$&HgBa$_2$CuO$_4$F&HgBa$_2$CuO$_5$\\
         \hline \hline
    $U_{\mathrm{bare}}$[eV]&12.278&13.131&10.639\\ 
        \hline
     $U$[eV]&2.946&1.544&1.055\\ 
    \hline
     $|U/t|$&6.547&3.088&2.045\\ 
     \hline
  \end{tabular}
\end{table} 

To discuss the reason why incommensurate FM fluctuations do not increase monotonically, we calculate an average of the single-particle spectral weight within $k_{\mathrm B}T$ around the Fermi level. The average spectral function is given as 
\begin{equation}
\tilde{z}(\bm k)\equiv-2G(\bm k,\tau=\beta/2)=\int^{\infty}_{-\infty}\frac{d\omega}{2\pi}\frac{A(\bm k,\omega)}{\cosh(\beta\omega/2)},
\end{equation}
where $A(\bm k,\omega)$ represents the spectral function~\cite{Vilk_1996,vilk1997non}, and the imaginary-time Green's function $G(\bm k,\tau)$ is obtained by $G(\bm k,\tau)=(1/\beta)\sum_{n}e^{-i\varepsilon_n\tau}G(\bm k,i\varepsilon_n)$. Figure~\ref{fig:z_U3_6_001_035_v2} shows the average spectral function for different strengths of $U$ in the overdoped case ($n=0.7$). One can see that the average spectral weights becomes blurred as $U$ increases, especially around "hot spots," i.e., $(\pi,0)$ and symmetrically equivalent momenta. This implies that the strength of FM spin fluctuation via $(q_x,q_y) \sim (0, 0)$ decreases with an increase in $U$. In contrast, incommensurate AFM spin fluctuations increase monotonically with an increase in $U$ because the average spectral weight along the node direction does not become blurred. These different self-energy effects between incommensurate FM and AFM fluctuations are directly reflected in the spectral weights. 

In conclusion, within constrained-RPA, we found that the on-site Coulomb interaction decreases with an increase in hole doping; however, the reduction in the interaction may be overestimated. We attribute the development of FM fluctuations in overdoped cuprates to the reduction in $U$. The filling dependency of the on-site Coulomb interaction is crucial to understand the emergence of FM fluctuations. 

Now, let us discuss the strength of $d$-wave superconductivity in accordance with the strength of the on-site Coulomb interaction for several fillings. In Fig.~\ref{fig:lambda}, we evaluate $\lambda$s at $k_{\mathrm B}T = 0.01$ eV. In optimal doping ($n=0.85$), $\lambda$s become larger than 1 in the range of $U \gtrsim 5$ eV. The results show that $d$-wave superconductivity is monotonically suppressed with increasing hole doping for several different values of $U$ due to the worse nesting condition~\cite{FLEX+DMFT}. In our scenario, the suppression of $d$-wave superconductivity arises not only from the worse band nesting but also from the reduction in on-site Coulomb interaction and, hence, the development of FM fluctuations.
\section{Summary and discussion}
In summary, we evaluated spin fluctuations in heavily overdoped cuprates. We found that the doping dependency of the on-site Coulomb interaction is one of the key factors that affect the strength of FM fluctuations. The strength of the on-site Coulomb interaction decreases with an increase in hole doping. Because high values of $U$ weaken FM fluctuations as a result of the self-energy effect, a good situation is realized owing to the reduction of $U$ for the development of FM fluctuations.

Although $d$-wave superconductivity can be suppressed with FM fluctuations in cuprate superconductors~\cite{cuprates_ruthenates_lindhard}, careful consideration is required for the quantitative assessment of superconductivity suppression. Indeed, some studies have suggested that FM fluctuations are not related to the disappearance of superconductivity in overdoped cuprates~\cite{wu2017}.

As a further research, FLEX with current vertex correction~\cite{kontani1999,arakawa2015,aritaCVC} may provide good experimental test for the characteristic temperature dependence of DC resistivity of metals with FM fluctuations in self-consistent renormalization theory of spin fluctuations. In addition, here, we did not pay considerable attention to material dependencies of cuprates; however, the model parameters (e.g., hopping parameters) should be optimized for each crystal structure. We should also carefully consider details of the doping dependence of crystal structures because we expect that the nesting vector $\bm q_2$ in some cases will be closer to (0,0) than in this study~\cite{RPA2015variousfillings,FM_TPSC_Tremblay}. In this study, we did not consider an effective model with Hund's rule coupling. Because Hund's rule coupling enhances FM fluctuations, the effect of this term in a multiband model can be relevant for quantitative discussion. This will be performed in future studies. Moreover, in the strong coupling regime, the Fermi surfaces for different doping rates are calculated within the slave-boson mean-field theory of the $t$-$J$ model~\cite{t-J_ogata}. The results show that Fermi surfaces can be modified only slightly using the $J$-term; a detailed study on this aspect will also be considered in the future. 

\begin{acknowledgments}
The calculations were performed in the computer centers of Kyushu University and ISSP, University of Tokyo. The authors would like to thank Enago (www.enago.jp) for the English langauage review.
\end{acknowledgments}
\bibliographystyle{apsrev4-2}
\bibliography{FMcuprates}

\begin{thebibliography}{45}%
\makeatletter
\providecommand \@ifxundefined [1]{%
 \@ifx{#1\undefined}
}%
\providecommand \@ifnum [1]{%
 \ifnum #1\expandafter \@firstoftwo
 \else \expandafter \@secondoftwo
 \fi
}%
\providecommand \@ifx [1]{%
 \ifx #1\expandafter \@firstoftwo
 \else \expandafter \@secondoftwo
 \fi
}%
\providecommand \natexlab [1]{#1}%
\providecommand \enquote  [1]{``#1''}%
\providecommand \bibnamefont  [1]{#1}%
\providecommand \bibfnamefont [1]{#1}%
\providecommand \citenamefont [1]{#1}%
\providecommand \href@noop [0]{\@secondoftwo}%
\providecommand \href [0]{\begingroup \@sanitize@url \@href}%
\providecommand \@href[1]{\@@startlink{#1}\@@href}%
\providecommand \@@href[1]{\endgroup#1\@@endlink}%
\providecommand \@sanitize@url [0]{\catcode `\\12\catcode `\$12\catcode
  `\&12\catcode `\#12\catcode `\^12\catcode `\_12\catcode `\%12\relax}%
\providecommand \@@startlink[1]{}%
\providecommand \@@endlink[0]{}%
\providecommand \url  [0]{\begingroup\@sanitize@url \@url }%
\providecommand \@url [1]{\endgroup\@href {#1}{\urlprefix }}%
\providecommand \urlprefix  [0]{URL }%
\providecommand \Eprint [0]{\href }%
\providecommand \doibase [0]{https://doi.org/}%
\providecommand \selectlanguage [0]{\@gobble}%
\providecommand \bibinfo  [0]{\@secondoftwo}%
\providecommand \bibfield  [0]{\@secondoftwo}%
\providecommand \translation [1]{[#1]}%
\providecommand \BibitemOpen [0]{}%
\providecommand \bibitemStop [0]{}%
\providecommand \bibitemNoStop [0]{.\EOS\space}%
\providecommand \EOS [0]{\spacefactor3000\relax}%
\providecommand \BibitemShut  [1]{\csname bibitem#1\endcsname}%
\let\auto@bib@innerbib\@empty
\bibitem [{\citenamefont {Kopp}\ \emph {et~al.}(2007)\citenamefont {Kopp},
  \citenamefont {Ghosal},\ and\ \citenamefont {Chakravarty}}]{FMkopp}%
  \BibitemOpen
  \bibfield  {author} {\bibinfo {author} {\bibfnamefont {A.}~\bibnamefont
  {Kopp}}, \bibinfo {author} {\bibfnamefont {A.}~\bibnamefont {Ghosal}},\ and\
  \bibinfo {author} {\bibfnamefont {S.}~\bibnamefont {Chakravarty}},\ }\href
  {https://doi.org/10.1073/pnas.0701265104} {\bibfield  {journal} {\bibinfo
  {journal} {Proceedings of the National Academy of Sciences}\ }\textbf
  {\bibinfo {volume} {104}},\ \bibinfo {pages} {6123} (\bibinfo {year}
  {2007})},\ \Eprint
  {https://arxiv.org/abs/https://www.pnas.org/content/104/15/6123.full.pdf}
  {https://www.pnas.org/content/104/15/6123.full.pdf} \BibitemShut {NoStop}%
\bibitem [{\citenamefont {Kurashima}\ \emph {et~al.}(2018)\citenamefont
  {Kurashima}, \citenamefont {Adachi}, \citenamefont {Suzuki}, \citenamefont
  {Fukunaga}, \citenamefont {Kawamata}, \citenamefont {Noji}, \citenamefont
  {Miyasaka}, \citenamefont {Watanabe}, \citenamefont {Miyazaki}, \citenamefont
  {Koda}, \citenamefont {Kadono},\ and\ \citenamefont {Koike}}]{FMkurashima}%
  \BibitemOpen
  \bibfield  {author} {\bibinfo {author} {\bibfnamefont {K.}~\bibnamefont
  {Kurashima}}, \bibinfo {author} {\bibfnamefont {T.}~\bibnamefont {Adachi}},
  \bibinfo {author} {\bibfnamefont {K.~M.}\ \bibnamefont {Suzuki}}, \bibinfo
  {author} {\bibfnamefont {Y.}~\bibnamefont {Fukunaga}}, \bibinfo {author}
  {\bibfnamefont {T.}~\bibnamefont {Kawamata}}, \bibinfo {author}
  {\bibfnamefont {T.}~\bibnamefont {Noji}}, \bibinfo {author} {\bibfnamefont
  {H.}~\bibnamefont {Miyasaka}}, \bibinfo {author} {\bibfnamefont
  {I.}~\bibnamefont {Watanabe}}, \bibinfo {author} {\bibfnamefont
  {M.}~\bibnamefont {Miyazaki}}, \bibinfo {author} {\bibfnamefont
  {A.}~\bibnamefont {Koda}}, \bibinfo {author} {\bibfnamefont {R.}~\bibnamefont
  {Kadono}},\ and\ \bibinfo {author} {\bibfnamefont {Y.}~\bibnamefont
  {Koike}},\ }\href {https://doi.org/10.1103/PhysRevLett.121.057002} {\bibfield
   {journal} {\bibinfo  {journal} {Phys. Rev. Lett.}\ }\textbf {\bibinfo
  {volume} {121}},\ \bibinfo {pages} {057002} (\bibinfo {year}
  {2018})}\BibitemShut {NoStop}%
\bibitem [{\citenamefont {Moriya}\ and\ \citenamefont
  {Kawabata}(1973{\natexlab{a}})}]{scr_1}%
  \BibitemOpen
  \bibfield  {author} {\bibinfo {author} {\bibfnamefont {T.}~\bibnamefont
  {Moriya}}\ and\ \bibinfo {author} {\bibfnamefont {A.}~\bibnamefont
  {Kawabata}},\ }\href {https://doi.org/10.1143/JPSJ.34.639} {\bibfield
  {journal} {\bibinfo  {journal} {Journal of the Physical Society of Japan}\
  }\textbf {\bibinfo {volume} {34}},\ \bibinfo {pages} {639} (\bibinfo {year}
  {1973}{\natexlab{a}})},\ \Eprint
  {https://arxiv.org/abs/https://doi.org/10.1143/JPSJ.34.639}
  {https://doi.org/10.1143/JPSJ.34.639} \BibitemShut {NoStop}%
\bibitem [{\citenamefont {Moriya}\ and\ \citenamefont
  {Kawabata}(1973{\natexlab{b}})}]{scr_2}%
  \BibitemOpen
  \bibfield  {author} {\bibinfo {author} {\bibfnamefont {T.}~\bibnamefont
  {Moriya}}\ and\ \bibinfo {author} {\bibfnamefont {A.}~\bibnamefont
  {Kawabata}},\ }\href {https://doi.org/10.1143/JPSJ.35.669} {\bibfield
  {journal} {\bibinfo  {journal} {Journal of the Physical Society of Japan}\
  }\textbf {\bibinfo {volume} {35}},\ \bibinfo {pages} {669} (\bibinfo {year}
  {1973}{\natexlab{b}})},\ \Eprint
  {https://arxiv.org/abs/https://doi.org/10.1143/JPSJ.35.669}
  {https://doi.org/10.1143/JPSJ.35.669} \BibitemShut {NoStop}%
\bibitem [{\citenamefont {Jia}\ \emph {et~al.}(2014)\citenamefont {Jia},
  \citenamefont {Nowadnick}, \citenamefont {Wohlfeld}, \citenamefont {Kung},
  \citenamefont {Chen}, \citenamefont {Johnston}, \citenamefont {Tohyama},
  \citenamefont {Moritz},\ and\ \citenamefont {Devereaux}}]{jia2014persistent}%
  \BibitemOpen
  \bibfield  {author} {\bibinfo {author} {\bibfnamefont {C.}~\bibnamefont
  {Jia}}, \bibinfo {author} {\bibfnamefont {E.}~\bibnamefont {Nowadnick}},
  \bibinfo {author} {\bibfnamefont {K.}~\bibnamefont {Wohlfeld}}, \bibinfo
  {author} {\bibfnamefont {Y.}~\bibnamefont {Kung}}, \bibinfo {author}
  {\bibfnamefont {C.-C.}\ \bibnamefont {Chen}}, \bibinfo {author}
  {\bibfnamefont {S.}~\bibnamefont {Johnston}}, \bibinfo {author}
  {\bibfnamefont {T.}~\bibnamefont {Tohyama}}, \bibinfo {author} {\bibfnamefont
  {B.}~\bibnamefont {Moritz}},\ and\ \bibinfo {author} {\bibfnamefont
  {T.}~\bibnamefont {Devereaux}},\ }\href@noop {} {\bibfield  {journal}
  {\bibinfo  {journal} {Nature communications}\ }\textbf {\bibinfo {volume}
  {5}},\ \bibinfo {pages} {1} (\bibinfo {year} {2014})}\BibitemShut {NoStop}%
\bibitem [{\citenamefont {Hankevych}\ \emph {et~al.}(2003)\citenamefont
  {Hankevych}, \citenamefont {Kyung},\ and\ \citenamefont
  {Tremblay}}]{FM_TPSC_Tremblay}%
  \BibitemOpen
  \bibfield  {author} {\bibinfo {author} {\bibfnamefont {V.}~\bibnamefont
  {Hankevych}}, \bibinfo {author} {\bibfnamefont {B.}~\bibnamefont {Kyung}},\
  and\ \bibinfo {author} {\bibfnamefont {A.-M.~S.}\ \bibnamefont {Tremblay}},\
  }\href {https://doi.org/10.1103/PhysRevB.68.214405} {\bibfield  {journal}
  {\bibinfo  {journal} {Phys. Rev. B}\ }\textbf {\bibinfo {volume} {68}},\
  \bibinfo {pages} {214405} (\bibinfo {year} {2003})}\BibitemShut {NoStop}%
\bibitem [{\citenamefont {Ogura}\ and\ \citenamefont
  {Kuroki}(2015)}]{OguraTPSC}%
  \BibitemOpen
  \bibfield  {author} {\bibinfo {author} {\bibfnamefont {D.}~\bibnamefont
  {Ogura}}\ and\ \bibinfo {author} {\bibfnamefont {K.}~\bibnamefont {Kuroki}},\
  }\href {https://doi.org/10.1103/PhysRevB.92.144511} {\bibfield  {journal}
  {\bibinfo  {journal} {Phys. Rev. B}\ }\textbf {\bibinfo {volume} {92}},\
  \bibinfo {pages} {144511} (\bibinfo {year} {2015})}\BibitemShut {NoStop}%
\bibitem [{\citenamefont {Maier}\ and\ \citenamefont
  {Scalapino}(2020)}]{maierscalapino2020}%
  \BibitemOpen
  \bibfield  {author} {\bibinfo {author} {\bibfnamefont {T.}~\bibnamefont
  {Maier}}\ and\ \bibinfo {author} {\bibfnamefont {D.}~\bibnamefont
  {Scalapino}},\ }\href {https://doi.org/10.1007/s10948-019-05366-4} {\bibfield
   {journal} {\bibinfo  {journal} {Journal of Superconductivity and Novel
  Magnetism}\ }\textbf {\bibinfo {volume} {33}} (\bibinfo {year}
  {2020})}\BibitemShut {NoStop}%
\bibitem [{\citenamefont {Maier}\ \emph {et~al.}(2020)\citenamefont {Maier},
  \citenamefont {Karakuzu},\ and\ \citenamefont
  {Scalapino}}]{PhysRevResearch.2.033132}%
  \BibitemOpen
  \bibfield  {author} {\bibinfo {author} {\bibfnamefont {T.~A.}\ \bibnamefont
  {Maier}}, \bibinfo {author} {\bibfnamefont {S.}~\bibnamefont {Karakuzu}},\
  and\ \bibinfo {author} {\bibfnamefont {D.~J.}\ \bibnamefont {Scalapino}},\
  }\href {https://doi.org/10.1103/PhysRevResearch.2.033132} {\bibfield
  {journal} {\bibinfo  {journal} {Phys. Rev. Research}\ }\textbf {\bibinfo
  {volume} {2}},\ \bibinfo {pages} {033132} (\bibinfo {year}
  {2020})}\BibitemShut {NoStop}%
\bibitem [{\citenamefont {Barbiellini}\ and\ \citenamefont
  {Jarlborg}(2008)}]{Barbiellini}%
  \BibitemOpen
  \bibfield  {author} {\bibinfo {author} {\bibfnamefont {B.}~\bibnamefont
  {Barbiellini}}\ and\ \bibinfo {author} {\bibfnamefont {T.}~\bibnamefont
  {Jarlborg}},\ }\href {https://doi.org/10.1103/PhysRevLett.101.157002}
  {\bibfield  {journal} {\bibinfo  {journal} {Phys. Rev. Lett.}\ }\textbf
  {\bibinfo {volume} {101}},\ \bibinfo {pages} {157002} (\bibinfo {year}
  {2008})}\BibitemShut {NoStop}%
\bibitem [{\citenamefont {Bickers}\ \emph {et~al.}(1989)\citenamefont
  {Bickers}, \citenamefont {Scalapino},\ and\ \citenamefont {White}}]{FLEX_1}%
  \BibitemOpen
  \bibfield  {author} {\bibinfo {author} {\bibfnamefont {N.~E.}\ \bibnamefont
  {Bickers}}, \bibinfo {author} {\bibfnamefont {D.~J.}\ \bibnamefont
  {Scalapino}},\ and\ \bibinfo {author} {\bibfnamefont {S.~R.}\ \bibnamefont
  {White}},\ }\href {https://doi.org/10.1103/PhysRevLett.62.961} {\bibfield
  {journal} {\bibinfo  {journal} {Phys. Rev. Lett.}\ }\textbf {\bibinfo
  {volume} {62}},\ \bibinfo {pages} {961} (\bibinfo {year} {1989})}\BibitemShut
  {NoStop}%
\bibitem [{\citenamefont {Bickers}\ and\ \citenamefont
  {Scalapino}(1989)}]{FLEX_2}%
  \BibitemOpen
  \bibfield  {author} {\bibinfo {author} {\bibfnamefont {N.}~\bibnamefont
  {Bickers}}\ and\ \bibinfo {author} {\bibfnamefont {D.}~\bibnamefont
  {Scalapino}},\ }\href
  {https://doi.org/https://doi.org/10.1016/0003-4916(89)90359-X} {\bibfield
  {journal} {\bibinfo  {journal} {Annals of Physics}\ }\textbf {\bibinfo
  {volume} {193}},\ \bibinfo {pages} {206 } (\bibinfo {year}
  {1989})}\BibitemShut {NoStop}%
\bibitem [{\citenamefont {Vishik}\ \emph {et~al.}(2014)\citenamefont {Vishik},
  \citenamefont {Bari\ifmmode \check{s}\else \v{s}\fi{}i\ifmmode~\acute{c}\else
  \'{c}\fi{}}, \citenamefont {Chan}, \citenamefont {Li}, \citenamefont {Xia},
  \citenamefont {Yu}, \citenamefont {Zhao}, \citenamefont {Lee}, \citenamefont
  {Meevasana}, \citenamefont {Devereaux}, \citenamefont {Greven},\ and\
  \citenamefont {Shen}}]{Hg1201_ARPES}%
  \BibitemOpen
  \bibfield  {author} {\bibinfo {author} {\bibfnamefont {I.~M.}\ \bibnamefont
  {Vishik}}, \bibinfo {author} {\bibfnamefont {N.}~\bibnamefont {Bari\ifmmode
  \check{s}\else \v{s}\fi{}i\ifmmode~\acute{c}\else \'{c}\fi{}}}, \bibinfo
  {author} {\bibfnamefont {M.~K.}\ \bibnamefont {Chan}}, \bibinfo {author}
  {\bibfnamefont {Y.}~\bibnamefont {Li}}, \bibinfo {author} {\bibfnamefont
  {D.~D.}\ \bibnamefont {Xia}}, \bibinfo {author} {\bibfnamefont
  {G.}~\bibnamefont {Yu}}, \bibinfo {author} {\bibfnamefont {X.}~\bibnamefont
  {Zhao}}, \bibinfo {author} {\bibfnamefont {W.~S.}\ \bibnamefont {Lee}},
  \bibinfo {author} {\bibfnamefont {W.}~\bibnamefont {Meevasana}}, \bibinfo
  {author} {\bibfnamefont {T.~P.}\ \bibnamefont {Devereaux}}, \bibinfo {author}
  {\bibfnamefont {M.}~\bibnamefont {Greven}},\ and\ \bibinfo {author}
  {\bibfnamefont {Z.-X.}\ \bibnamefont {Shen}},\ }\href
  {https://doi.org/10.1103/PhysRevB.89.195141} {\bibfield  {journal} {\bibinfo
  {journal} {Phys. Rev. B}\ }\textbf {\bibinfo {volume} {89}},\ \bibinfo
  {pages} {195141} (\bibinfo {year} {2014})}\BibitemShut {NoStop}%
\bibitem [{\citenamefont {Lehmann}\ and\ \citenamefont
  {Taut}(1972)}]{tetra_origin_1}%
  \BibitemOpen
  \bibfield  {author} {\bibinfo {author} {\bibfnamefont {G.}~\bibnamefont
  {Lehmann}}\ and\ \bibinfo {author} {\bibfnamefont {M.}~\bibnamefont {Taut}},\
  }\href {https://doi.org/10.1002/pssb.2220540211} {\bibfield  {journal}
  {\bibinfo  {journal} {physica status solidi (b)}\ }\textbf {\bibinfo {volume}
  {54}},\ \bibinfo {pages} {469} (\bibinfo {year} {1972})},\ \Eprint
  {https://arxiv.org/abs/https://onlinelibrary.wiley.com/doi/pdf/10.1002/pssb.2220540211}
  {https://onlinelibrary.wiley.com/doi/pdf/10.1002/pssb.2220540211}
  \BibitemShut {NoStop}%
\bibitem [{\citenamefont {Rath}\ and\ \citenamefont
  {Freeman}(1975)}]{tetra_origin_2}%
  \BibitemOpen
  \bibfield  {author} {\bibinfo {author} {\bibfnamefont {J.}~\bibnamefont
  {Rath}}\ and\ \bibinfo {author} {\bibfnamefont {A.~J.}\ \bibnamefont
  {Freeman}},\ }\href {https://doi.org/10.1103/PhysRevB.11.2109} {\bibfield
  {journal} {\bibinfo  {journal} {Phys. Rev. B}\ }\textbf {\bibinfo {volume}
  {11}},\ \bibinfo {pages} {2109} (\bibinfo {year} {1975})}\BibitemShut
  {NoStop}%
\bibitem [{\citenamefont {Putilin}\ \emph {et~al.}(1993)\citenamefont
  {Putilin}, \citenamefont {Antipov}, \citenamefont {Chmaissem},\ and\
  \citenamefont {Marezio}}]{Hg1201-first}%
  \BibitemOpen
  \bibfield  {author} {\bibinfo {author} {\bibfnamefont {S.}~\bibnamefont
  {Putilin}}, \bibinfo {author} {\bibfnamefont {E.}~\bibnamefont {Antipov}},
  \bibinfo {author} {\bibfnamefont {O.}~\bibnamefont {Chmaissem}},\ and\
  \bibinfo {author} {\bibfnamefont {M.}~\bibnamefont {Marezio}},\ }\href@noop
  {} {\bibfield  {journal} {\bibinfo  {journal} {Nature}\ }\textbf {\bibinfo
  {volume} {362}},\ \bibinfo {pages} {226} (\bibinfo {year}
  {1993})}\BibitemShut {NoStop}%
\bibitem [{\citenamefont {Franchini}\ \emph {et~al.}(2000)\citenamefont
  {Franchini}, \citenamefont {Massidda}, \citenamefont {Continenza},\ and\
  \citenamefont {Gauzzi}}]{Hg1201_F_DFT}%
  \BibitemOpen
  \bibfield  {author} {\bibinfo {author} {\bibfnamefont {C.}~\bibnamefont
  {Franchini}}, \bibinfo {author} {\bibfnamefont {S.}~\bibnamefont {Massidda}},
  \bibinfo {author} {\bibfnamefont {A.}~\bibnamefont {Continenza}},\ and\
  \bibinfo {author} {\bibfnamefont {A.}~\bibnamefont {Gauzzi}},\ }\href
  {https://doi.org/10.1103/PhysRevB.62.9163} {\bibfield  {journal} {\bibinfo
  {journal} {Phys. Rev. B}\ }\textbf {\bibinfo {volume} {62}},\ \bibinfo
  {pages} {9163} (\bibinfo {year} {2000})}\BibitemShut {NoStop}%
\bibitem [{\citenamefont {Antipov}\ \emph {et~al.}(1997)\citenamefont
  {Antipov}, \citenamefont {Abakumov}, \citenamefont {Aksenov}, \citenamefont
  {Balagurov}, \citenamefont {Putilin},\ and\ \citenamefont
  {Rozova}}]{Hg1201_F_neutron}%
  \BibitemOpen
  \bibfield  {author} {\bibinfo {author} {\bibfnamefont {E.}~\bibnamefont
  {Antipov}}, \bibinfo {author} {\bibfnamefont {A.}~\bibnamefont {Abakumov}},
  \bibinfo {author} {\bibfnamefont {V.}~\bibnamefont {Aksenov}}, \bibinfo
  {author} {\bibfnamefont {A.}~\bibnamefont {Balagurov}}, \bibinfo {author}
  {\bibfnamefont {S.}~\bibnamefont {Putilin}},\ and\ \bibinfo {author}
  {\bibfnamefont {M.}~\bibnamefont {Rozova}},\ }\href
  {https://doi.org/https://doi.org/10.1016/S0921-4526(97)00715-1} {\bibfield
  {journal} {\bibinfo  {journal} {Physica B: Condensed Matter}\ }\textbf
  {\bibinfo {volume} {241-243}},\ \bibinfo {pages} {773 } (\bibinfo {year}
  {1997})},\ \bibinfo {note} {proceedings of the International Conference on
  Neutron Scattering}\BibitemShut {NoStop}%
\bibitem [{\citenamefont {Agrawal}\ and\ \citenamefont
  {Agrawal}(1994)}]{HgBa2CuO5_DFT}%
  \BibitemOpen
  \bibfield  {author} {\bibinfo {author} {\bibfnamefont {B.~K.}\ \bibnamefont
  {Agrawal}}\ and\ \bibinfo {author} {\bibfnamefont {S.}~\bibnamefont
  {Agrawal}},\ }\href
  {https://doi.org/https://doi.org/10.1016/0921-4534(94)00177-4} {\bibfield
  {journal} {\bibinfo  {journal} {Physica C: Superconductivity}\ }\textbf
  {\bibinfo {volume} {233}},\ \bibinfo {pages} {8 } (\bibinfo {year}
  {1994})}\BibitemShut {NoStop}%
\bibitem [{\citenamefont {Sakakibara}\ \emph {et~al.}(2012)\citenamefont
  {Sakakibara}, \citenamefont {Usui}, \citenamefont {Kuroki}, \citenamefont
  {Arita},\ and\ \citenamefont {Aoki}}]{Sakakibara2012}%
  \BibitemOpen
  \bibfield  {author} {\bibinfo {author} {\bibfnamefont {H.}~\bibnamefont
  {Sakakibara}}, \bibinfo {author} {\bibfnamefont {H.}~\bibnamefont {Usui}},
  \bibinfo {author} {\bibfnamefont {K.}~\bibnamefont {Kuroki}}, \bibinfo
  {author} {\bibfnamefont {R.}~\bibnamefont {Arita}},\ and\ \bibinfo {author}
  {\bibfnamefont {H.}~\bibnamefont {Aoki}},\ }\href
  {https://doi.org/10.1103/PhysRevB.85.064501} {\bibfield  {journal} {\bibinfo
  {journal} {Phys. Rev. B}\ }\textbf {\bibinfo {volume} {85}},\ \bibinfo
  {pages} {1} (\bibinfo {year} {2012})},\ \Eprint
  {https://arxiv.org/abs/1112.0964} {arXiv:1112.0964} \BibitemShut {NoStop}%
\bibitem [{\citenamefont {Giannozzi}\ \emph {et~al.}(2009)\citenamefont
  {Giannozzi}, \citenamefont {Baroni}, \citenamefont {Bonini}, \citenamefont
  {Calandra}, \citenamefont {Car}, \citenamefont {Cavazzoni}, \citenamefont
  {Ceresoli}, \citenamefont {Chiarotti}, \citenamefont {Cococcioni},
  \citenamefont {Dabo}, \citenamefont {{Dal Corso}}, \citenamefont
  {de~Gironcoli}, \citenamefont {Fabris}, \citenamefont {Fratesi},
  \citenamefont {Gebauer}, \citenamefont {Gerstmann}, \citenamefont
  {Gougoussis}, \citenamefont {Kokalj}, \citenamefont {Lazzeri}, \citenamefont
  {Martin-Samos}, \citenamefont {Marzari}, \citenamefont {Mauri}, \citenamefont
  {Mazzarello}, \citenamefont {Paolini}, \citenamefont {Pasquarello},
  \citenamefont {Paulatto}, \citenamefont {Sbraccia}, \citenamefont {Scandolo},
  \citenamefont {Sclauzero}, \citenamefont {Seitsonen}, \citenamefont
  {Smogunov}, \citenamefont {Umari},\ and\ \citenamefont
  {Wentzcovitch}}]{QE-2009}%
  \BibitemOpen
  \bibfield  {author} {\bibinfo {author} {\bibfnamefont {P.}~\bibnamefont
  {Giannozzi}}, \bibinfo {author} {\bibfnamefont {S.}~\bibnamefont {Baroni}},
  \bibinfo {author} {\bibfnamefont {N.}~\bibnamefont {Bonini}}, \bibinfo
  {author} {\bibfnamefont {M.}~\bibnamefont {Calandra}}, \bibinfo {author}
  {\bibfnamefont {R.}~\bibnamefont {Car}}, \bibinfo {author} {\bibfnamefont
  {C.}~\bibnamefont {Cavazzoni}}, \bibinfo {author} {\bibfnamefont
  {D.}~\bibnamefont {Ceresoli}}, \bibinfo {author} {\bibfnamefont {G.~L.}\
  \bibnamefont {Chiarotti}}, \bibinfo {author} {\bibfnamefont {M.}~\bibnamefont
  {Cococcioni}}, \bibinfo {author} {\bibfnamefont {I.}~\bibnamefont {Dabo}},
  \bibinfo {author} {\bibfnamefont {A.}~\bibnamefont {{Dal Corso}}}, \bibinfo
  {author} {\bibfnamefont {S.}~\bibnamefont {de~Gironcoli}}, \bibinfo {author}
  {\bibfnamefont {S.}~\bibnamefont {Fabris}}, \bibinfo {author} {\bibfnamefont
  {G.}~\bibnamefont {Fratesi}}, \bibinfo {author} {\bibfnamefont
  {R.}~\bibnamefont {Gebauer}}, \bibinfo {author} {\bibfnamefont
  {U.}~\bibnamefont {Gerstmann}}, \bibinfo {author} {\bibfnamefont
  {C.}~\bibnamefont {Gougoussis}}, \bibinfo {author} {\bibfnamefont
  {A.}~\bibnamefont {Kokalj}}, \bibinfo {author} {\bibfnamefont
  {M.}~\bibnamefont {Lazzeri}}, \bibinfo {author} {\bibfnamefont
  {L.}~\bibnamefont {Martin-Samos}}, \bibinfo {author} {\bibfnamefont
  {N.}~\bibnamefont {Marzari}}, \bibinfo {author} {\bibfnamefont
  {F.}~\bibnamefont {Mauri}}, \bibinfo {author} {\bibfnamefont
  {R.}~\bibnamefont {Mazzarello}}, \bibinfo {author} {\bibfnamefont
  {S.}~\bibnamefont {Paolini}}, \bibinfo {author} {\bibfnamefont
  {A.}~\bibnamefont {Pasquarello}}, \bibinfo {author} {\bibfnamefont
  {L.}~\bibnamefont {Paulatto}}, \bibinfo {author} {\bibfnamefont
  {C.}~\bibnamefont {Sbraccia}}, \bibinfo {author} {\bibfnamefont
  {S.}~\bibnamefont {Scandolo}}, \bibinfo {author} {\bibfnamefont
  {G.}~\bibnamefont {Sclauzero}}, \bibinfo {author} {\bibfnamefont {A.~P.}\
  \bibnamefont {Seitsonen}}, \bibinfo {author} {\bibfnamefont {A.}~\bibnamefont
  {Smogunov}}, \bibinfo {author} {\bibfnamefont {P.}~\bibnamefont {Umari}},\
  and\ \bibinfo {author} {\bibfnamefont {R.~M.}\ \bibnamefont {Wentzcovitch}},\
  }\href {http://www.quantum-espresso.org} {\bibfield  {journal} {\bibinfo
  {journal} {Journal of Physics: Condensed Matter}\ }\textbf {\bibinfo {volume}
  {21}},\ \bibinfo {pages} {395502 (19pp)} (\bibinfo {year}
  {2009})}\BibitemShut {NoStop}%
\bibitem [{\citenamefont {Giannozzi}\ \emph {et~al.}(2017)\citenamefont
  {Giannozzi}, \citenamefont {Andreussi}, \citenamefont {Brumme}, \citenamefont
  {Bunau}, \citenamefont {Nardelli}, \citenamefont {Calandra}, \citenamefont
  {Car}, \citenamefont {Cavazzoni}, \citenamefont {Ceresoli}, \citenamefont
  {Cococcioni}, \citenamefont {Colonna}, \citenamefont {Carnimeo},
  \citenamefont {Corso}, \citenamefont {de~Gironcoli}, \citenamefont {Delugas},
  \citenamefont {Jr}, \citenamefont {Ferretti}, \citenamefont {Floris},
  \citenamefont {Fratesi}, \citenamefont {Fugallo}, \citenamefont {Gebauer},
  \citenamefont {Gerstmann}, \citenamefont {Giustino}, \citenamefont {Gorni},
  \citenamefont {Jia}, \citenamefont {Kawamura}, \citenamefont {Ko},
  \citenamefont {Kokalj}, \citenamefont {K\"{u}\c{c}\"{u}kbenli}, \citenamefont
  {Lazzeri}, \citenamefont {Marsili}, \citenamefont {Marzari}, \citenamefont
  {Mauri}, \citenamefont {Nguyen}, \citenamefont {Nguyen}, \citenamefont {de-la
  Roza}, \citenamefont {Paulatto}, \citenamefont {Ponc\'e}, \citenamefont
  {Rocca}, \citenamefont {Sabatini}, \citenamefont {Santra}, \citenamefont
  {Schlipf}, \citenamefont {Seitsonen}, \citenamefont {Smogunov}, \citenamefont
  {Timrov}, \citenamefont {Thonhauser}, \citenamefont {Umari}, \citenamefont
  {Vast}, \citenamefont {Wu},\ and\ \citenamefont {Baroni}}]{QE-2017}%
  \BibitemOpen
  \bibfield  {author} {\bibinfo {author} {\bibfnamefont {P.}~\bibnamefont
  {Giannozzi}}, \bibinfo {author} {\bibfnamefont {O.}~\bibnamefont
  {Andreussi}}, \bibinfo {author} {\bibfnamefont {T.}~\bibnamefont {Brumme}},
  \bibinfo {author} {\bibfnamefont {O.}~\bibnamefont {Bunau}}, \bibinfo
  {author} {\bibfnamefont {M.~B.}\ \bibnamefont {Nardelli}}, \bibinfo {author}
  {\bibfnamefont {M.}~\bibnamefont {Calandra}}, \bibinfo {author}
  {\bibfnamefont {R.}~\bibnamefont {Car}}, \bibinfo {author} {\bibfnamefont
  {C.}~\bibnamefont {Cavazzoni}}, \bibinfo {author} {\bibfnamefont
  {D.}~\bibnamefont {Ceresoli}}, \bibinfo {author} {\bibfnamefont
  {M.}~\bibnamefont {Cococcioni}}, \bibinfo {author} {\bibfnamefont
  {N.}~\bibnamefont {Colonna}}, \bibinfo {author} {\bibfnamefont
  {I.}~\bibnamefont {Carnimeo}}, \bibinfo {author} {\bibfnamefont {A.~D.}\
  \bibnamefont {Corso}}, \bibinfo {author} {\bibfnamefont {S.}~\bibnamefont
  {de~Gironcoli}}, \bibinfo {author} {\bibfnamefont {P.}~\bibnamefont
  {Delugas}}, \bibinfo {author} {\bibfnamefont {R.~A.~D.}\ \bibnamefont {Jr}},
  \bibinfo {author} {\bibfnamefont {A.}~\bibnamefont {Ferretti}}, \bibinfo
  {author} {\bibfnamefont {A.}~\bibnamefont {Floris}}, \bibinfo {author}
  {\bibfnamefont {G.}~\bibnamefont {Fratesi}}, \bibinfo {author} {\bibfnamefont
  {G.}~\bibnamefont {Fugallo}}, \bibinfo {author} {\bibfnamefont
  {R.}~\bibnamefont {Gebauer}}, \bibinfo {author} {\bibfnamefont
  {U.}~\bibnamefont {Gerstmann}}, \bibinfo {author} {\bibfnamefont
  {F.}~\bibnamefont {Giustino}}, \bibinfo {author} {\bibfnamefont
  {T.}~\bibnamefont {Gorni}}, \bibinfo {author} {\bibfnamefont
  {J.}~\bibnamefont {Jia}}, \bibinfo {author} {\bibfnamefont {M.}~\bibnamefont
  {Kawamura}}, \bibinfo {author} {\bibfnamefont {H.-Y.}\ \bibnamefont {Ko}},
  \bibinfo {author} {\bibfnamefont {A.}~\bibnamefont {Kokalj}}, \bibinfo
  {author} {\bibfnamefont {E.}~\bibnamefont {K\"{u}\c{c}\"{u}kbenli}}, \bibinfo
  {author} {\bibfnamefont {M.}~\bibnamefont {Lazzeri}}, \bibinfo {author}
  {\bibfnamefont {M.}~\bibnamefont {Marsili}}, \bibinfo {author} {\bibfnamefont
  {N.}~\bibnamefont {Marzari}}, \bibinfo {author} {\bibfnamefont
  {F.}~\bibnamefont {Mauri}}, \bibinfo {author} {\bibfnamefont {N.~L.}\
  \bibnamefont {Nguyen}}, \bibinfo {author} {\bibfnamefont {H.-V.}\
  \bibnamefont {Nguyen}}, \bibinfo {author} {\bibfnamefont {A.~O.}\
  \bibnamefont {de-la Roza}}, \bibinfo {author} {\bibfnamefont
  {L.}~\bibnamefont {Paulatto}}, \bibinfo {author} {\bibfnamefont
  {S.}~\bibnamefont {Ponc\'e}}, \bibinfo {author} {\bibfnamefont
  {D.}~\bibnamefont {Rocca}}, \bibinfo {author} {\bibfnamefont
  {R.}~\bibnamefont {Sabatini}}, \bibinfo {author} {\bibfnamefont
  {B.}~\bibnamefont {Santra}}, \bibinfo {author} {\bibfnamefont
  {M.}~\bibnamefont {Schlipf}}, \bibinfo {author} {\bibfnamefont {A.~P.}\
  \bibnamefont {Seitsonen}}, \bibinfo {author} {\bibfnamefont {A.}~\bibnamefont
  {Smogunov}}, \bibinfo {author} {\bibfnamefont {I.}~\bibnamefont {Timrov}},
  \bibinfo {author} {\bibfnamefont {T.}~\bibnamefont {Thonhauser}}, \bibinfo
  {author} {\bibfnamefont {P.}~\bibnamefont {Umari}}, \bibinfo {author}
  {\bibfnamefont {N.}~\bibnamefont {Vast}}, \bibinfo {author} {\bibfnamefont
  {X.}~\bibnamefont {Wu}},\ and\ \bibinfo {author} {\bibfnamefont
  {S.}~\bibnamefont {Baroni}},\ }\href
  {http://stacks.iop.org/0953-8984/29/i=46/a=465901} {\bibfield  {journal}
  {\bibinfo  {journal} {Journal of Physics: Condensed Matter}\ }\textbf
  {\bibinfo {volume} {29}},\ \bibinfo {pages} {465901} (\bibinfo {year}
  {2017})}\BibitemShut {NoStop}%
\bibitem [{\citenamefont {Souza}\ \emph {et~al.}(2001)\citenamefont {Souza},
  \citenamefont {Marzari},\ and\ \citenamefont {Vanderbilt}}]{wannier_ent}%
  \BibitemOpen
  \bibfield  {author} {\bibinfo {author} {\bibfnamefont {I.}~\bibnamefont
  {Souza}}, \bibinfo {author} {\bibfnamefont {N.}~\bibnamefont {Marzari}},\
  and\ \bibinfo {author} {\bibfnamefont {D.}~\bibnamefont {Vanderbilt}},\
  }\href {https://doi.org/10.1103/PhysRevB.65.035109} {\bibfield  {journal}
  {\bibinfo  {journal} {Phys. Rev. B}\ }\textbf {\bibinfo {volume} {65}},\
  \bibinfo {pages} {035109} (\bibinfo {year} {2001})}\BibitemShut {NoStop}%
\bibitem [{\citenamefont {Marzari}\ and\ \citenamefont
  {Vanderbilt}(1997)}]{wannier_loc}%
  \BibitemOpen
  \bibfield  {author} {\bibinfo {author} {\bibfnamefont {N.}~\bibnamefont
  {Marzari}}\ and\ \bibinfo {author} {\bibfnamefont {D.}~\bibnamefont
  {Vanderbilt}},\ }\href {https://doi.org/10.1103/PhysRevB.56.12847} {\bibfield
   {journal} {\bibinfo  {journal} {Phys. Rev. B}\ }\textbf {\bibinfo {volume}
  {56}},\ \bibinfo {pages} {12847} (\bibinfo {year} {1997})}\BibitemShut
  {NoStop}%
\bibitem [{\citenamefont {Mostofi}\ \emph {et~al.}(2014)\citenamefont
  {Mostofi}, \citenamefont {Yates}, \citenamefont {Pizzi}, \citenamefont {Lee},
  \citenamefont {Souza}, \citenamefont {Vanderbilt},\ and\ \citenamefont
  {Marzari}}]{wannier90}%
  \BibitemOpen
  \bibfield  {author} {\bibinfo {author} {\bibfnamefont {A.~A.}\ \bibnamefont
  {Mostofi}}, \bibinfo {author} {\bibfnamefont {J.~R.}\ \bibnamefont {Yates}},
  \bibinfo {author} {\bibfnamefont {G.}~\bibnamefont {Pizzi}}, \bibinfo
  {author} {\bibfnamefont {Y.-S.}\ \bibnamefont {Lee}}, \bibinfo {author}
  {\bibfnamefont {I.}~\bibnamefont {Souza}}, \bibinfo {author} {\bibfnamefont
  {D.}~\bibnamefont {Vanderbilt}},\ and\ \bibinfo {author} {\bibfnamefont
  {N.}~\bibnamefont {Marzari}},\ }\href
  {https://doi.org/https://doi.org/10.1016/j.cpc.2014.05.003} {\bibfield
  {journal} {\bibinfo  {journal} {Computer Physics Communications}\ }\textbf
  {\bibinfo {volume} {185}},\ \bibinfo {pages} {2309 } (\bibinfo {year}
  {2014})}\BibitemShut {NoStop}%
\bibitem [{\citenamefont {Aryasetiawan}\ \emph {et~al.}(2004)\citenamefont
  {Aryasetiawan}, \citenamefont {Imada}, \citenamefont {Georges}, \citenamefont
  {Kotliar}, \citenamefont {Biermann},\ and\ \citenamefont
  {Lichtenstein}}]{cRPA}%
  \BibitemOpen
  \bibfield  {author} {\bibinfo {author} {\bibfnamefont {F.}~\bibnamefont
  {Aryasetiawan}}, \bibinfo {author} {\bibfnamefont {M.}~\bibnamefont {Imada}},
  \bibinfo {author} {\bibfnamefont {A.}~\bibnamefont {Georges}}, \bibinfo
  {author} {\bibfnamefont {G.}~\bibnamefont {Kotliar}}, \bibinfo {author}
  {\bibfnamefont {S.}~\bibnamefont {Biermann}},\ and\ \bibinfo {author}
  {\bibfnamefont {A.~I.}\ \bibnamefont {Lichtenstein}},\ }\href
  {https://doi.org/10.1103/PhysRevB.70.195104} {\bibfield  {journal} {\bibinfo
  {journal} {Phys. Rev. B}\ }\textbf {\bibinfo {volume} {70}},\ \bibinfo
  {pages} {195104} (\bibinfo {year} {2004})}\BibitemShut {NoStop}%
\bibitem [{\citenamefont {Aryasetiawan}\ \emph {et~al.}(2006)\citenamefont
  {Aryasetiawan}, \citenamefont {Karlsson}, \citenamefont {Jepsen},\ and\
  \citenamefont {Sch\"onberger}}]{c-rpa_pd2}%
  \BibitemOpen
  \bibfield  {author} {\bibinfo {author} {\bibfnamefont {F.}~\bibnamefont
  {Aryasetiawan}}, \bibinfo {author} {\bibfnamefont {K.}~\bibnamefont
  {Karlsson}}, \bibinfo {author} {\bibfnamefont {O.}~\bibnamefont {Jepsen}},\
  and\ \bibinfo {author} {\bibfnamefont {U.}~\bibnamefont {Sch\"onberger}},\
  }\href {https://doi.org/10.1103/PhysRevB.74.125106} {\bibfield  {journal}
  {\bibinfo  {journal} {Phys. Rev. B}\ }\textbf {\bibinfo {volume} {74}},\
  \bibinfo {pages} {125106} (\bibinfo {year} {2006})}\BibitemShut {NoStop}%
\bibitem [{\citenamefont {Miyake}\ and\ \citenamefont
  {Aryasetiawan}(2008)}]{c-rpa_pd1}%
  \BibitemOpen
  \bibfield  {author} {\bibinfo {author} {\bibfnamefont {T.}~\bibnamefont
  {Miyake}}\ and\ \bibinfo {author} {\bibfnamefont {F.}~\bibnamefont
  {Aryasetiawan}},\ }\href {https://doi.org/10.1103/PhysRevB.77.085122}
  {\bibfield  {journal} {\bibinfo  {journal} {Phys. Rev. B}\ }\textbf {\bibinfo
  {volume} {77}},\ \bibinfo {pages} {085122} (\bibinfo {year}
  {2008})}\BibitemShut {NoStop}%
\bibitem [{\citenamefont {Miyake}\ \emph {et~al.}(2009)\citenamefont {Miyake},
  \citenamefont {Aryasetiawan},\ and\ \citenamefont {Imada}}]{c-rpa_pd3}%
  \BibitemOpen
  \bibfield  {author} {\bibinfo {author} {\bibfnamefont {T.}~\bibnamefont
  {Miyake}}, \bibinfo {author} {\bibfnamefont {F.}~\bibnamefont
  {Aryasetiawan}},\ and\ \bibinfo {author} {\bibfnamefont {M.}~\bibnamefont
  {Imada}},\ }\href {https://doi.org/10.1103/PhysRevB.80.155134} {\bibfield
  {journal} {\bibinfo  {journal} {Phys. Rev. B}\ }\textbf {\bibinfo {volume}
  {80}},\ \bibinfo {pages} {155134} (\bibinfo {year} {2009})}\BibitemShut
  {NoStop}%
\bibitem [{\citenamefont {Nakamura}\ \emph {et~al.}(2016)\citenamefont
  {Nakamura}, \citenamefont {Nohara}, \citenamefont {Yoshimoto},\ and\
  \citenamefont {Nomura}}]{respack1}%
  \BibitemOpen
  \bibfield  {author} {\bibinfo {author} {\bibfnamefont {K.}~\bibnamefont
  {Nakamura}}, \bibinfo {author} {\bibfnamefont {Y.}~\bibnamefont {Nohara}},
  \bibinfo {author} {\bibfnamefont {Y.}~\bibnamefont {Yoshimoto}},\ and\
  \bibinfo {author} {\bibfnamefont {Y.}~\bibnamefont {Nomura}},\ }\href
  {https://doi.org/10.1103/PhysRevB.93.085124} {\bibfield  {journal} {\bibinfo
  {journal} {Phys. Rev. B}\ }\textbf {\bibinfo {volume} {93}},\ \bibinfo
  {pages} {085124} (\bibinfo {year} {2016})}\BibitemShut {NoStop}%
\bibitem [{\citenamefont {Nakamura}\ \emph {et~al.}(2009)\citenamefont
  {Nakamura}, \citenamefont {Yoshimoto}, \citenamefont {Kosugi}, \citenamefont
  {Arita},\ and\ \citenamefont {Imada}}]{respack2}%
  \BibitemOpen
  \bibfield  {author} {\bibinfo {author} {\bibfnamefont {K.}~\bibnamefont
  {Nakamura}}, \bibinfo {author} {\bibfnamefont {Y.}~\bibnamefont {Yoshimoto}},
  \bibinfo {author} {\bibfnamefont {T.}~\bibnamefont {Kosugi}}, \bibinfo
  {author} {\bibfnamefont {R.}~\bibnamefont {Arita}},\ and\ \bibinfo {author}
  {\bibfnamefont {M.}~\bibnamefont {Imada}},\ }\href
  {https://doi.org/10.1143/JPSJ.78.083710} {\bibfield  {journal} {\bibinfo
  {journal} {J. Phys. Soc. Jpn}\ }\textbf {\bibinfo {volume} {78}},\ \bibinfo
  {pages} {083710} (\bibinfo {year} {2009})},\ \Eprint
  {https://arxiv.org/abs/https://doi.org/10.1143/JPSJ.78.083710}
  {https://doi.org/10.1143/JPSJ.78.083710} \BibitemShut {NoStop}%
\bibitem [{\citenamefont {Nakamura}\ \emph {et~al.}(2008)\citenamefont
  {Nakamura}, \citenamefont {Arita},\ and\ \citenamefont {Imada}}]{respack3}%
  \BibitemOpen
  \bibfield  {author} {\bibinfo {author} {\bibfnamefont {K.}~\bibnamefont
  {Nakamura}}, \bibinfo {author} {\bibfnamefont {R.}~\bibnamefont {Arita}},\
  and\ \bibinfo {author} {\bibfnamefont {M.}~\bibnamefont {Imada}},\ }\href
  {https://doi.org/10.1143/JPSJ.77.093711} {\bibfield  {journal} {\bibinfo
  {journal} {J. Phys. Soc. Jpn}\ }\textbf {\bibinfo {volume} {77}},\ \bibinfo
  {pages} {093711} (\bibinfo {year} {2008})},\ \Eprint
  {https://arxiv.org/abs/https://doi.org/10.1143/JPSJ.77.093711}
  {https://doi.org/10.1143/JPSJ.77.093711} \BibitemShut {NoStop}%
\bibitem [{\citenamefont {Nohara}\ \emph {et~al.}(2009)\citenamefont {Nohara},
  \citenamefont {Yamamoto},\ and\ \citenamefont {Fujiwara}}]{respack4}%
  \BibitemOpen
  \bibfield  {author} {\bibinfo {author} {\bibfnamefont {Y.}~\bibnamefont
  {Nohara}}, \bibinfo {author} {\bibfnamefont {S.}~\bibnamefont {Yamamoto}},\
  and\ \bibinfo {author} {\bibfnamefont {T.}~\bibnamefont {Fujiwara}},\ }\href
  {https://doi.org/10.1103/PhysRevB.79.195110} {\bibfield  {journal} {\bibinfo
  {journal} {Phys. Rev. B}\ }\textbf {\bibinfo {volume} {79}},\ \bibinfo
  {pages} {195110} (\bibinfo {year} {2009})}\BibitemShut {NoStop}%
\bibitem [{\citenamefont {Fujiwara}\ \emph {et~al.}(2003)\citenamefont
  {Fujiwara}, \citenamefont {Yamamoto},\ and\ \citenamefont
  {Ishii}}]{respack5}%
  \BibitemOpen
  \bibfield  {author} {\bibinfo {author} {\bibfnamefont {T.}~\bibnamefont
  {Fujiwara}}, \bibinfo {author} {\bibfnamefont {S.}~\bibnamefont {Yamamoto}},\
  and\ \bibinfo {author} {\bibfnamefont {Y.}~\bibnamefont {Ishii}},\ }\href
  {https://doi.org/10.1143/JPSJ.72.777} {\bibfield  {journal} {\bibinfo
  {journal} {J. Phys. Soc. Jpn}\ }\textbf {\bibinfo {volume} {72}},\ \bibinfo
  {pages} {777} (\bibinfo {year} {2003})},\ \Eprint
  {https://arxiv.org/abs/https://doi.org/10.1143/JPSJ.72.777}
  {https://doi.org/10.1143/JPSJ.72.777} \BibitemShut {NoStop}%
\bibitem [{\citenamefont {Teranishi}\ \emph {et~al.}(2018)\citenamefont
  {Teranishi}, \citenamefont {Nishiguchi},\ and\ \citenamefont
  {Kusakabe}}]{cRPAteranishi}%
  \BibitemOpen
  \bibfield  {author} {\bibinfo {author} {\bibfnamefont {S.}~\bibnamefont
  {Teranishi}}, \bibinfo {author} {\bibfnamefont {K.}~\bibnamefont
  {Nishiguchi}},\ and\ \bibinfo {author} {\bibfnamefont {K.}~\bibnamefont
  {Kusakabe}},\ }\href {https://doi.org/10.7566/JPSJ.87.114701} {\bibfield
  {journal} {\bibinfo  {journal} {Journal of the Physical Society of Japan}\
  }\textbf {\bibinfo {volume} {87}},\ \bibinfo {pages} {114701} (\bibinfo
  {year} {2018})},\ \Eprint
  {https://arxiv.org/abs/https://doi.org/10.7566/JPSJ.87.114701}
  {https://doi.org/10.7566/JPSJ.87.114701} \BibitemShut {NoStop}%
\bibitem [{\citenamefont {Vilk}\ and\ \citenamefont
  {Tremblay}(1996)}]{Vilk_1996}%
  \BibitemOpen
  \bibfield  {author} {\bibinfo {author} {\bibfnamefont {Y.~M.}\ \bibnamefont
  {Vilk}}\ and\ \bibinfo {author} {\bibfnamefont {A.-M.~S.}\ \bibnamefont
  {Tremblay}},\ }\href {https://doi.org/10.1209/epl/i1996-00315-2} {\bibfield
  {journal} {\bibinfo  {journal} {Europhysics Letters ({EPL})}\ }\textbf
  {\bibinfo {volume} {33}},\ \bibinfo {pages} {159} (\bibinfo {year}
  {1996})}\BibitemShut {NoStop}%
\bibitem [{\citenamefont {Vilk}\ and\ \citenamefont
  {Tremblay}(1997)}]{vilk1997non}%
  \BibitemOpen
  \bibfield  {author} {\bibinfo {author} {\bibfnamefont {Y.}~\bibnamefont
  {Vilk}}\ and\ \bibinfo {author} {\bibfnamefont {A.-M.}\ \bibnamefont
  {Tremblay}},\ }\href@noop {} {\bibfield  {journal} {\bibinfo  {journal}
  {Journal de Physique I}\ }\textbf {\bibinfo {volume} {7}},\ \bibinfo {pages}
  {1309} (\bibinfo {year} {1997})}\BibitemShut {NoStop}%
\bibitem [{\citenamefont {Kitatani}\ \emph {et~al.}(2015)\citenamefont
  {Kitatani}, \citenamefont {Tsuji},\ and\ \citenamefont {Aoki}}]{FLEX+DMFT}%
  \BibitemOpen
  \bibfield  {author} {\bibinfo {author} {\bibfnamefont {M.}~\bibnamefont
  {Kitatani}}, \bibinfo {author} {\bibfnamefont {N.}~\bibnamefont {Tsuji}},\
  and\ \bibinfo {author} {\bibfnamefont {H.}~\bibnamefont {Aoki}},\ }\href
  {https://doi.org/10.1103/PhysRevB.92.085104} {\bibfield  {journal} {\bibinfo
  {journal} {Phys. Rev. B}\ }\textbf {\bibinfo {volume} {92}},\ \bibinfo
  {pages} {085104} (\bibinfo {year} {2015})}\BibitemShut {NoStop}%
\bibitem [{\citenamefont {Monthoux}\ and\ \citenamefont
  {Lonzarich}(2005)}]{cuprates_ruthenates_lindhard}%
  \BibitemOpen
  \bibfield  {author} {\bibinfo {author} {\bibfnamefont {P.}~\bibnamefont
  {Monthoux}}\ and\ \bibinfo {author} {\bibfnamefont {G.~G.}\ \bibnamefont
  {Lonzarich}},\ }\href {https://doi.org/10.1103/PhysRevB.71.054504} {\bibfield
   {journal} {\bibinfo  {journal} {Phys. Rev. B}\ }\textbf {\bibinfo {volume}
  {71}},\ \bibinfo {pages} {054504} (\bibinfo {year} {2005})}\BibitemShut
  {NoStop}%
\bibitem [{\citenamefont {Wu}\ \emph {et~al.}(2017)\citenamefont {Wu},
  \citenamefont {Lauter}, \citenamefont {Ambaye}, \citenamefont {He},\ and\
  \citenamefont {Bo{\v{z}}ovi{\'c}}}]{wu2017}%
  \BibitemOpen
  \bibfield  {author} {\bibinfo {author} {\bibfnamefont {J.}~\bibnamefont
  {Wu}}, \bibinfo {author} {\bibfnamefont {V.}~\bibnamefont {Lauter}}, \bibinfo
  {author} {\bibfnamefont {H.}~\bibnamefont {Ambaye}}, \bibinfo {author}
  {\bibfnamefont {X.}~\bibnamefont {He}},\ and\ \bibinfo {author}
  {\bibfnamefont {I.}~\bibnamefont {Bo{\v{z}}ovi{\'c}}},\ }\href@noop {}
  {\bibfield  {journal} {\bibinfo  {journal} {Scientific reports}\ }\textbf
  {\bibinfo {volume} {7}},\ \bibinfo {pages} {45896} (\bibinfo {year}
  {2017})}\BibitemShut {NoStop}%
\bibitem [{\citenamefont {Kontani}\ \emph {et~al.}(1999)\citenamefont
  {Kontani}, \citenamefont {Kanki},\ and\ \citenamefont {Ueda}}]{kontani1999}%
  \BibitemOpen
  \bibfield  {author} {\bibinfo {author} {\bibfnamefont {H.}~\bibnamefont
  {Kontani}}, \bibinfo {author} {\bibfnamefont {K.}~\bibnamefont {Kanki}},\
  and\ \bibinfo {author} {\bibfnamefont {K.}~\bibnamefont {Ueda}},\ }\href
  {https://doi.org/10.1103/PhysRevB.59.14723} {\bibfield  {journal} {\bibinfo
  {journal} {Phys. Rev. B}\ }\textbf {\bibinfo {volume} {59}},\ \bibinfo
  {pages} {14723} (\bibinfo {year} {1999})}\BibitemShut {NoStop}%
\bibitem [{\citenamefont {Arakawa}(2014)}]{arakawa2015}%
  \BibitemOpen
  \bibfield  {author} {\bibinfo {author} {\bibfnamefont {N.}~\bibnamefont
  {Arakawa}},\ }\href {https://doi.org/10.1103/PhysRevB.90.245103} {\bibfield
  {journal} {\bibinfo  {journal} {Phys. Rev. B}\ }\textbf {\bibinfo {volume}
  {90}},\ \bibinfo {pages} {245103} (\bibinfo {year} {2014})}\BibitemShut
  {NoStop}%
\bibitem [{\citenamefont {Arita}\ \emph {et~al.}(2000)\citenamefont {Arita},
  \citenamefont {Kuroki},\ and\ \citenamefont {Aoki}}]{aritaCVC}%
  \BibitemOpen
  \bibfield  {author} {\bibinfo {author} {\bibfnamefont {R.}~\bibnamefont
  {Arita}}, \bibinfo {author} {\bibfnamefont {K.}~\bibnamefont {Kuroki}},\ and\
  \bibinfo {author} {\bibfnamefont {H.}~\bibnamefont {Aoki}},\ }\href
  {https://doi.org/10.1103/PhysRevB.61.3207} {\bibfield  {journal} {\bibinfo
  {journal} {Phys. Rev. B}\ }\textbf {\bibinfo {volume} {61}},\ \bibinfo
  {pages} {3207} (\bibinfo {year} {2000})}\BibitemShut {NoStop}%
\bibitem [{\citenamefont {R\o{}mer}\ \emph {et~al.}(2015)\citenamefont
  {R\o{}mer}, \citenamefont {Kreisel}, \citenamefont {Eremin}, \citenamefont
  {Malakhov}, \citenamefont {Maier}, \citenamefont {Hirschfeld},\ and\
  \citenamefont {Andersen}}]{RPA2015variousfillings}%
  \BibitemOpen
  \bibfield  {author} {\bibinfo {author} {\bibfnamefont {A.~T.}\ \bibnamefont
  {R\o{}mer}}, \bibinfo {author} {\bibfnamefont {A.}~\bibnamefont {Kreisel}},
  \bibinfo {author} {\bibfnamefont {I.}~\bibnamefont {Eremin}}, \bibinfo
  {author} {\bibfnamefont {M.~A.}\ \bibnamefont {Malakhov}}, \bibinfo {author}
  {\bibfnamefont {T.~A.}\ \bibnamefont {Maier}}, \bibinfo {author}
  {\bibfnamefont {P.~J.}\ \bibnamefont {Hirschfeld}},\ and\ \bibinfo {author}
  {\bibfnamefont {B.~M.}\ \bibnamefont {Andersen}},\ }\href
  {https://doi.org/10.1103/PhysRevB.92.104505} {\bibfield  {journal} {\bibinfo
  {journal} {Phys. Rev. B}\ }\textbf {\bibinfo {volume} {92}},\ \bibinfo
  {pages} {104505} (\bibinfo {year} {2015})}\BibitemShut {NoStop}%
\bibitem [{\citenamefont {Ogata}\ and\ \citenamefont
  {Fukuyama}(2008)}]{t-J_ogata}%
  \BibitemOpen
  \bibfield  {author} {\bibinfo {author} {\bibfnamefont {M.}~\bibnamefont
  {Ogata}}\ and\ \bibinfo {author} {\bibfnamefont {H.}~\bibnamefont
  {Fukuyama}},\ }\href {http://stacks.iop.org/0034-4885/71/i=3/a=036501}
  {\bibfield  {journal} {\bibinfo  {journal} {Reports on Progress in Physics}\
  }\textbf {\bibinfo {volume} {71}},\ \bibinfo {pages} {036501} (\bibinfo
  {year} {2008})}\BibitemShut {NoStop}%
\end{thebibliography}%
\end{document}